\documentclass[prd,superscriptaddress,twocolumn]{revtex4-1}[11pt] 
\pdfoutput=1
\usepackage{amsmath,amssymb,graphicx} 
\usepackage{hyperref}
\usepackage{color}

\begin{document}

\title{Two distinct components of the delayed single electron background signals in liquid xenon emission detectors}
\author{P. Sorensen} \email{pfsorensen@lbl.gov}
\affiliation{Lawrence Berkeley National Laboratory, 1 Cyclotron Rd., Berkeley, CA 94720, USA} 

\author{K. Kamdin}
\affiliation{Lawrence Berkeley National Laboratory, 1 Cyclotron Rd., Berkeley, CA 94720, USA} 
\affiliation{University of California Berkeley, Department of Physics, Berkeley, CA 94720, USA} 
 
\begin{abstract}
Single electron background signals with millisecond timescales are known to follow ionizing events in liquid/gas xenon emission detectors. Due to the long timescale, these signals can present a limiting background to the low-energy threshold of dark matter searches, and prevent discovery-class searches for MeV scale hidden sector dark matter. A systematic study reveals a fast ($\tau_1$) and slow ($\tau_2$) component to the background. The fast component is compatible with electrons which were trapped at the liquid surface, and can be reduced by increasing the electric field. However, the slow component increases linearly with electric field. Hypotheses for the origin of the effect are discussed, and techniques for mitigation are suggested.
\end{abstract}
\date{\today}

\maketitle


\section{{Introduction}}\label{sec:intro}
Liquid xenon emission detectors \cite{Bolozdynya:1995,Aprile:2009dv,Bolozdynya:2010} have proven extremely useful for rare-event searches, such as for galactic dark matter. They are a variant of the time projection chamber in which ionized electrons are drifted to the liquid-gas interface under the influence of an applied electric field $E_l$. There, the electrons are extracted, drifted through the gas and amplified via proportional scintillation. This technique can easily provide sensitivity to single electrons. It has been known for decades that the bulk of the electron emission from the liquid into the gas proceeds in a time $t\lesssim1$~ns, and that there also exists a field-dependent component of the emission with $\tau \gtrsim0.1$~ms  \cite{Gushchin:1982a}. A widely accepted interpretation is that this  is due to thermalization of electrons at the interface \cite{Bolozdynya:1999}. 

In a search for low-mass dark matter, the XENON10 Collaboration identified a single electron background which they referred to as an ``electron train,'' with a time scale at least as long as the putative thermalized component \cite{Angle:2011th}. This background is thought to have been responsible for a significant number of the one, two and three electron signals reported in that analysis. A similar background was observed in ZEPLIN-III \cite{Santos:2011ju}, there-in referred to as `spontaneous' single electron signals, due to the lack of apparent time correlation with preceding events. More recently, dedicated studies \cite{Akimov:2016,Xu:2016} confirmed that indeed the spontaneous signals are time-correlated, and continue for many miliseconds after the primary event. 


There is great interest in mitigating this background. It is a potential irritation in the analysis of dark matter search data from instruments such as LUX \cite{Akerib:2016vxi} or XENON1T \cite{Aprile:2017iyp}, where it can complicate the classification of low-energy events. It greatly diminishes the discovery potential of these detectors in low-mass dark matter searches \cite{Angle:2011th,Aprile:2016wwo}, or MeV-scale hidden sector dark matter searches \cite{Essig:2011nj,Essig:2012yx,Essig:2017kqs}. Recognizing this, the proposed $U_{A'}(1)$ experiment \cite{Battaglieri:2017aum} aspires to deploy a 10-kg scale liquid xenon emission detector, free from electron train backgrounds. 

In this article we report results of R\&D towards that end, from a systematic study of electron trains in a small liquid xenon emission detector. The data clearly show two components $\tau_1$ and $\tau_2$ in the electron train, the faster of which ($\tau_1$) can reasonably be interpreted as the initially un-emitted (thermalized) electrons. However, the slower component ($\tau_2$) is not at all consistent with this hypothesis. Instead, it may be due to fluorescence photons which then photo-ionize impurities in the liquid xenon. 

\section{Experimental Details}
The test bed used to obtain these results was configured with a 50~mm diameter wire grid cathode and planar, segmented anode. The stainless steel wires had 100~$\mu$m diameter and were spaced on a 2~mm pitch. The anode and the cathode were located 5~mm apart, and $\sim650$~g of liquid xenon were introduced, until the liquid level was about 4~mm above the cathode. An electric field $E_l$ was applied in the $z$ direction by biasing the cathode to a voltage $-V$, and holding the anode at ground. Electrons resulting from ionizing radiation were thus drifted toward the anode, and extracted across the liquid-gas interface. The value of $E_l$ contains a systematic uncertainty of $\pm0.4$~kV/cm due to uncertainty in the liquid xenon level. This relationship is described in the Appendix. This uncertainty does not affect the basic results, but should be born in mind when comparing against other measurements. Aqueous $^{210}$Po (as PoCl$_4$) was evaporated on a single section of a single wire, in the center of the cathode. The literature does not discuss spontaneous deposition of polonium onto stainless steel \cite{Figgins:1981}, however, we found it to be robust against dissolution in liquid xenon. The source spot size was such that alpha particle signals only appeared on the charge preamplifier connected to the central, 6~mm diameter anode pad. This arrangement was made in order to mitigate the effects of electric field fringing near the edges of the active region. 

A single Hamamatsu R8778 VUV-sensitive photomultiplier was installed 10~mm below the cathode grid. The photomultiplier was biased to the recommended maximum -1500~V, resulting in easily recognizable single photoelectron pulses and a gain of $1.9\times10^7$.  Voltage records of ionizing radiation events were digitized at 14~bits with 125~MHz sampling and a 20~MHz low-pass filter, with an event duration of 1~ms. Typical events contained both a primary scintillation pulse (S1) and a secondary, proportional scintillation pulse (S2) caused by extracted electrons. The S2 signal was verified to be extremely linear, with the number of detected photons increasing as $y=10*(V-1.9)$. A helium after-pulse (a photomultiplier artifact) was often evident 500~ns after the S1.  The 5.3 MeV alpha particles result in an average 380,000 quanta (electrons and photons) in the liquid xenon, of which an electric-field-dependent $91.3\%$ (lowest $E_l$ data point) to $81.6\%$ (highest $E_l$ data point) appeared as $7.1$~eV primary scintillation photons, with the remainder as ionized  electrons. These signals were sufficient to saturate the biasing circuit of the photomultiplier, precluding an accurate measurement of the number of electrons present in the alpha S2. Despite this, (a) the alpha particle population was easily identifiable by its size and characteristic drift time (from the cathode), and (b) the number of electrons obtained from alpha particle pulses was measured separately with the charge-sensitive preamplifier.

During operation, the xenon was maintained at a vapor pressure of 1.5~Bar, with a cryostat temperature of 174~K. It was continuously circulated through a heated getter. The liquid was evaporated into the purification loop via a capillary, and the gas was separately purged into the purification loop via a sintered metal filter (to limit the purge rate). The xenon turnover time was about 6 hours, as measured by mass flow. About 2/3 of the flow rate was from the liquid, and the rest from the gas purge. The electron attenuation length was not measured during this experiment, but from comparison with previous experiments and measured charge yields in the literature, can be inferred to have improved to a few tens of cm over a week.


\begin{figure}[h]
\begin{center}
\vskip -0.0cm
\includegraphics[width=0.49\textwidth]{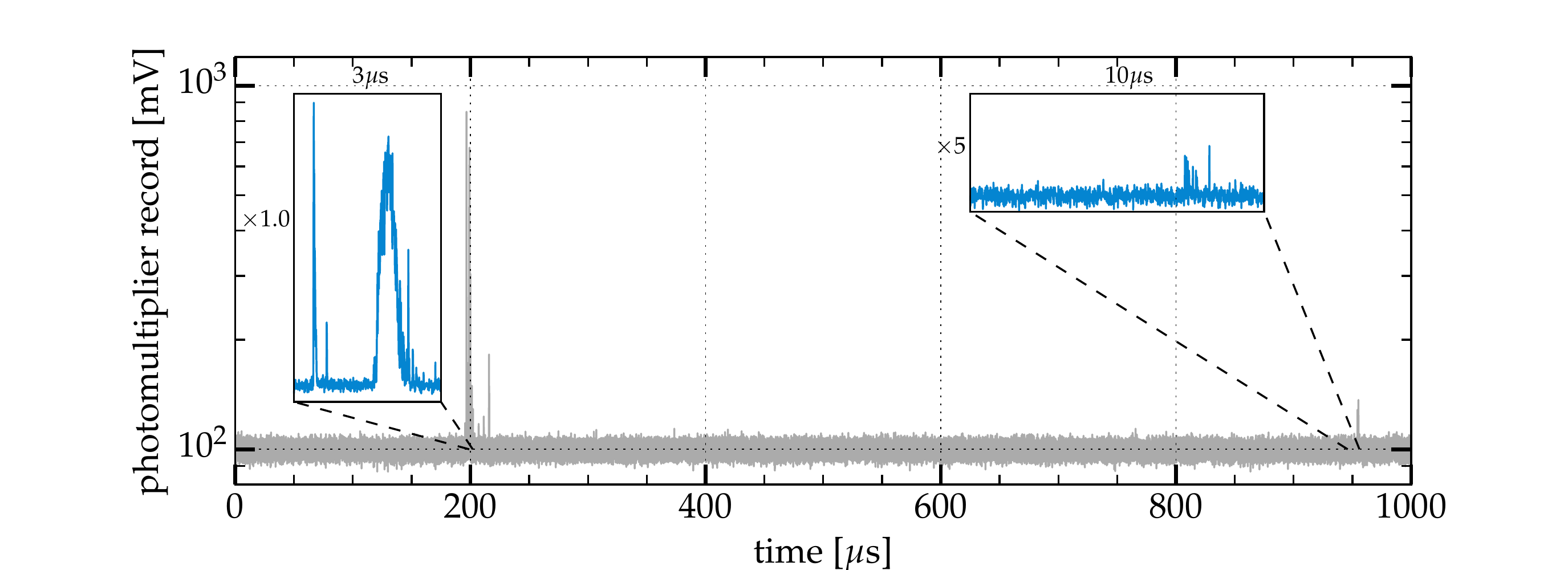}
\includegraphics[width=0.49\textwidth]{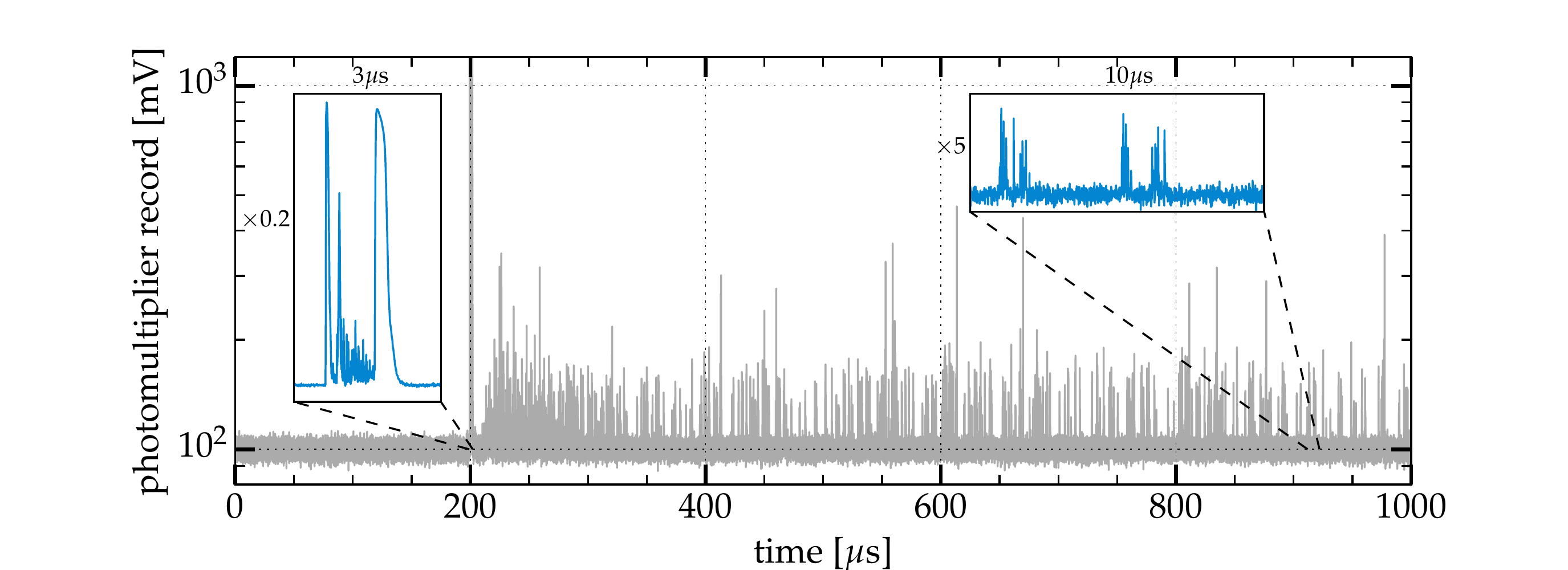}
\includegraphics[width=0.49\textwidth]{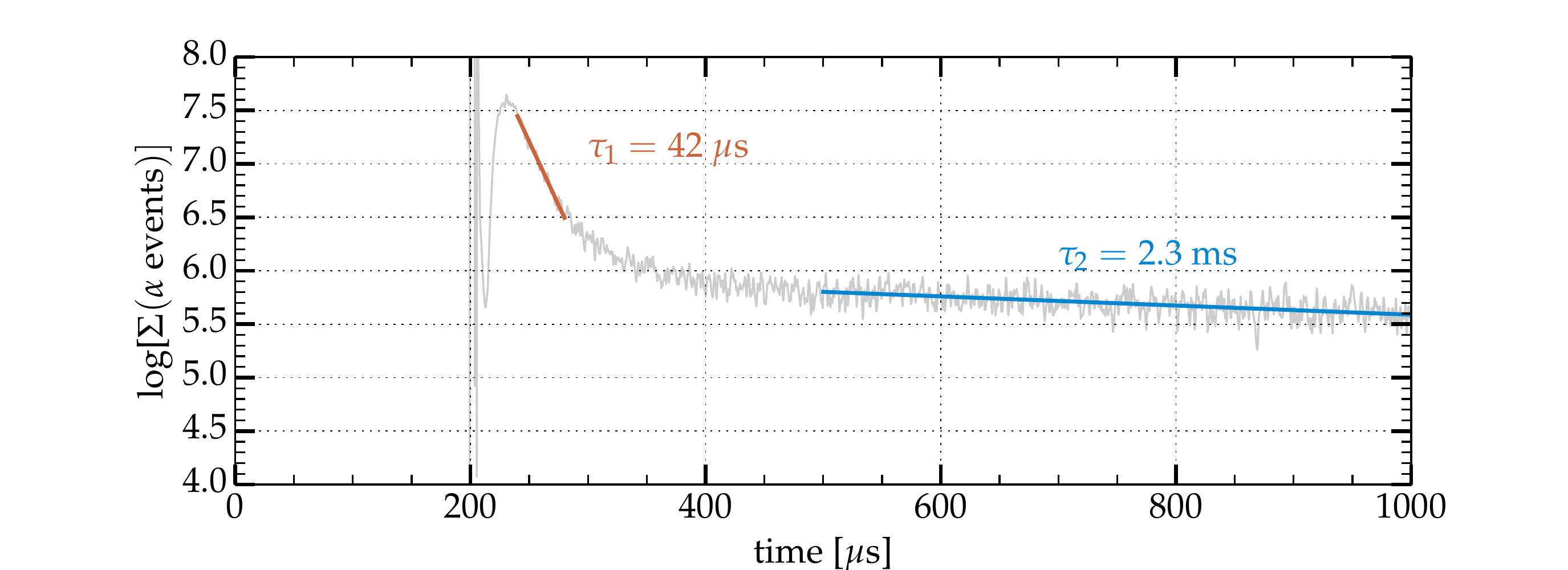}
   \vskip -0.1cm
\caption{Typical $^{206}$Pb and $^{210}$Po alpha particle events with $V=4$~kV, showing inset detail of the primary S1 and S2 structure of the event, as well as a zoom on a portion of the after-pulse region. {\bf (top)} A $^{206}$Pb recoil event record, with S1=130 detected photons and S2=67 electrons. A single electron is visible nearly 800~$\mu$s after the primary event. {\bf (middle)} A $^{210}$Po alpha particle event record. {\bf (bottom)} 400 summed alpha particle event records, low-pass filtered to more clearly illustrate the electron train.}
\vskip -0.5cm
\label{fig:events}
\end{center}
\end{figure} 


\section{Results and Discussion}
Data were acquired at four values of $E_l$, corresponding to $V=[3,4,5,6]$~kV.  In each case, the population of $^{210}$Po alpha particle events was isolated using simple, robust software cuts on the signal size of S1, S2 and the time delay between them.  The waveforms were summed and the decay time of the electron train was obtained by fitting a linear function to $y = \mbox{log}[\Sigma(\alpha~\mbox{events})]$ over 40~$\mu$s ($\tau_1$) and 500~$\mu$s ($\tau_2$) regions, indicated by the extent of the fits in Fig. \ref{fig:events} (bottom). The $\tau_1$ fit region was chosen to begin $40~\mu$s after the S2, to ensure fidelity of the photomultiplier signal. The number of single electrons in these regions were counted. Separately, the number of primary (prompt) electrons was measured with the charge sensitive preamplifier. 

\begin{figure}[th]
\begin{center}
\vskip -0.0cm
\includegraphics[width=0.49\textwidth]{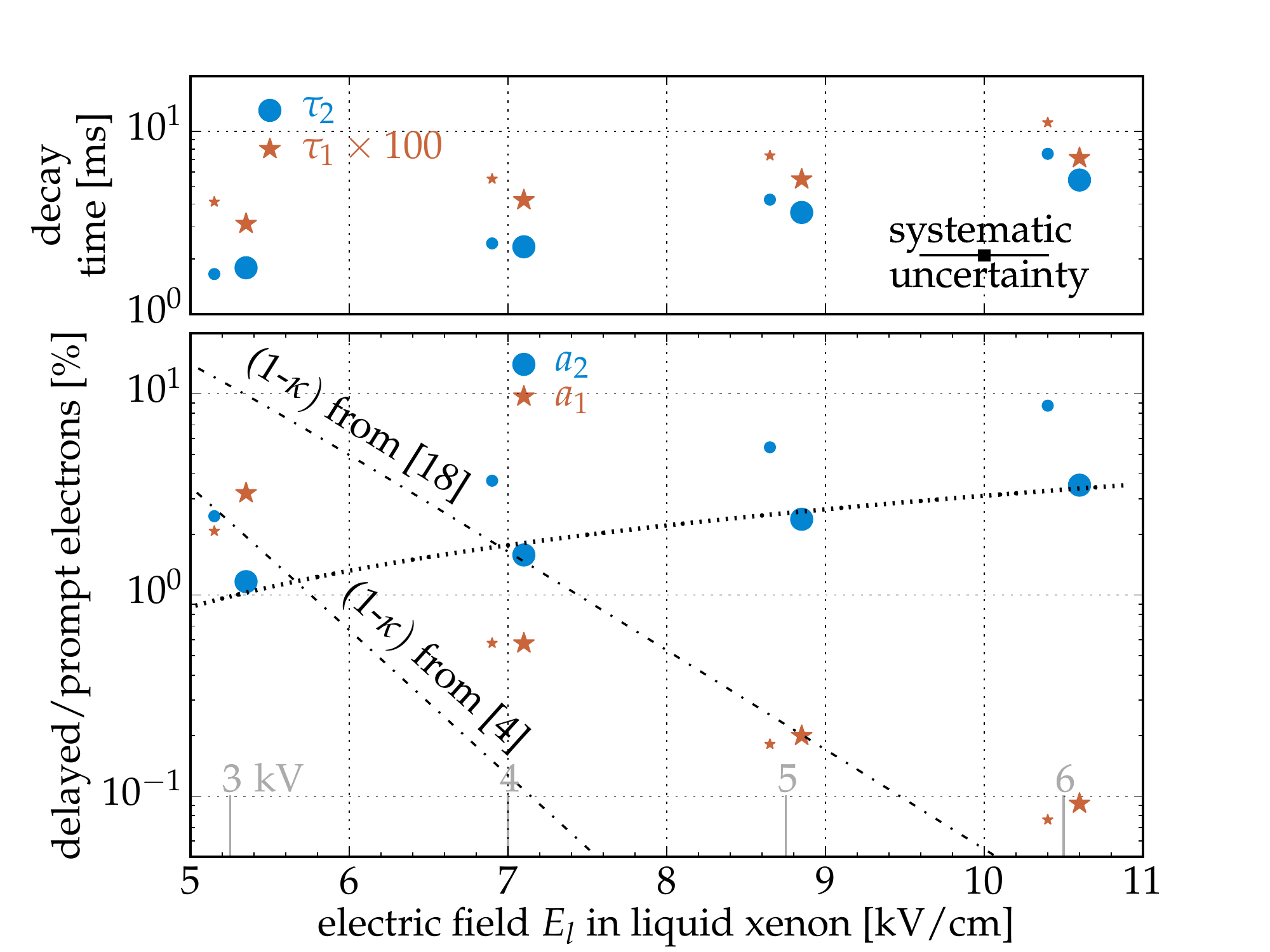}
\vskip -0.1cm
\caption{For each cathode potential (indicated along the $x$ axis), two data points are shown. The smaller points (offset to the left) were obtained while liquid xenon purity was relatively poor compared with the larger points (offset to the right). Systematic uncertainty in $E_l$ applies to all points. Statistical uncertainty is comparable to the size of the points. {\bf (top)} The measured decay times $\tau_1$ and $\tau_2$.  {\bf (bottom)} The amplitude of the electron train as a percent of the original number of electrons in the ionizing event, as a function of $E_l$. The fast component $a_1$ can be compared against $(1-\kappa)$, where $\kappa$ is the electron emission efficiency. The slow component $a_2$ is fit by a linear function (dotted). }
\vskip -0.5cm
\label{fig:percent}
\end{center}
\end{figure} 


The fast ($\tau_1$) and slow ($\tau_2$) components were assumed to begin immediately after the primary event, and extend to infinity. The measured amplitudes $a_1$ and $a_2$ of each component were corrected for the sample window, and are plotted in Fig. \ref{fig:percent} as a percent of the measured number of primary electrons. Two data sets are plotted for each value of $E_l$: data points offset to the left of each $E_l$ were acquired shortly after condensing xenon into the test bed, and data points offset to the right of each $E_l$ were acquired a week later, when the liquid purity had improved.

The fast component $a_1$ decreases with increasing $E_l$, suggesting that these are the thermalized, un-emitted primary electrons. This interpretation can be compared against $(1-\kappa)$, where $\kappa$ is the electron emission efficiency (expressed as a percent).  The parameterizations $\kappa~=~[1~+~\mbox{exp}(\frac{5}{3}(-E_l+3))]^{-1}$ and $\kappa~=~[1~+~\mbox{exp}(1.14*(-E_l+3.41))]^{-1}$ were fit to the data from \cite{Gushchin:1982a,Edwards:2017emx}. Assuming the measurements of \cite{Edwards:2017emx} to be correct, it would  appear that for $E_l<7$~kV/cm about half of the thermalized electrons are never emitted. This is consistent with capture by electronegative impurities before they can escape into the gas.

On the other hand, the slow component $a_2$ increases linearly with increasing $E_l$. This suggests an origin due to the number proportional scintillation photons, or due to the increase in electric field. A distinct reduction in the size of the slow component of the electron train is obtained by purifying the xenon. This suggests that at times $t\gtrsim 5\tau_1$, the single electrons in the electron train are due to ionization of impurities.

We also observe an increased rate of isolated single photons amongst the delayed electron signal. Some fraction of these photons are due to after-pulsing, a photomultiplier tube artifact. Controlling for this and the photomultiplier tube dark counts, delayed single photons seem to be present at a rate of about $\times0.1$ compared with the number of single electrons. However, because this test bed has no $(x,y)$ resolution in it's scintillation detection, we cannot exclude the possibility that some of these single photons might be due to single electrons whose proportional scintillation yield was very low (such as near the edges of the active region).

An immediate hypothesis for the origin of this effect is that Teflon, which comprises the walls of the detector, is weakly fluorescent following exposure to ultraviolet photons. Teflon has previously been seen to fluoresce in the visible region; the decay time is not known \cite{Gachkovskii:1970,Khatipov:2011}. The fluorescence photons would then photo-ionize the impurities.   It is also possible that either the xenon itself, or impurities within the xenon, are responsible for the fluorescence. It has been known for at least half a century \cite{Jortner:1965} that liquid xenon scintillates in the vacuum ultraviolet, due to formation of excimers with $\sim1$~eV binding energy. Later work categorized (e.g.) the decay times of the singlet $^1\Sigma_u^+$ and triplet $^3\Sigma_u^+$ states \cite{Kubota:1978}, which are a few to a few tens of nanoseconds. Data are sparse concerning bound states in heavy noble gas excimers, but additional higher-energy states are expected to be weakly bound \cite{Ermler:1978}. These states would naturally be excited by the process of proportional scintillation. Again, the lifetimes are not known. 



A final hypothesis is that the electron train is due to electron release from electronegative impurities. Impurities are well-known to capture electrons from the primary ionization event. The resulting negative ion then drifts slowly to the liquid-gas interface. It could release it's captured electron via collision, or tunneling. The former method of release is expected to have a timescale of about 10 seconds, which should not increase with increasing $E_l$. On the other hand, the timescale for tunneling would likely decrease with $E_l$. 

We note that care must be taken with the interpretation of the measured decay constants $\tau_1$ and $\tau_2$. In the limit of a static xenon fluid with perfect purity, $\tau_1$ would be the thermalized electron emission time. In actuality, the value of $\tau_1$ is likely shortened by the capture time of electrons by electronegative impurities. The value of $\tau_2$ may be affected by a number of different variables, depending on which hypothesis is correct. For example, if Teflon is responsible for the fluorescence, it's preparation and purity likely influence $\tau_2$. If molecules in the xenon are responsible for the fluorescence, their flow velocity and the optical properties of the detector influence $\tau_2$.

Finally, it is worth reminding the reader that additional components may exist amongst the delayed electron pulses. These could possibly be resolved with longer-timescale studies. A test bed capable of imaging the $(x,y)$ location of single electrons might also return additional information. And, other classes of delayed electron backgrounds have been noted \cite{Xu:2016}, which are not studied in the present work.

\section{Conclusions}
Delayed, time-correlated single electrons, also known as ``electron train'' backgrounds, in liquid xenon emission detectors show two distinct components. The fast component decreases with $E_l$ and appears to be due to thermalized electrons which were not emitted in the primary $S2$ signal. The slow component increases with $E_l$ and is consistent with several hypotheses, as discussed in the preceding section.


Previous suggestions for mitigation \cite{Sorensen:2017ymt} are relevant to the $\tau_1$ component, and therefore would not be sufficient for informing the design of discovery-class liquid xenon emission detectors for MeV-scale hidden sector dark matter. The present work suggests that regardless of the origin of the $\tau_2$ component, an immediate steps towards mitigation of this background is to improve the xenon purity.  A future work will attempt to identify the exact origin of this effect, by (1) removing Teflon from the detector construction, and (2) investigating the effect of impurity dopants.
 
 \begin{figure}[h]
\begin{center}
\vskip -0.0cm
\includegraphics[width=0.49\textwidth]{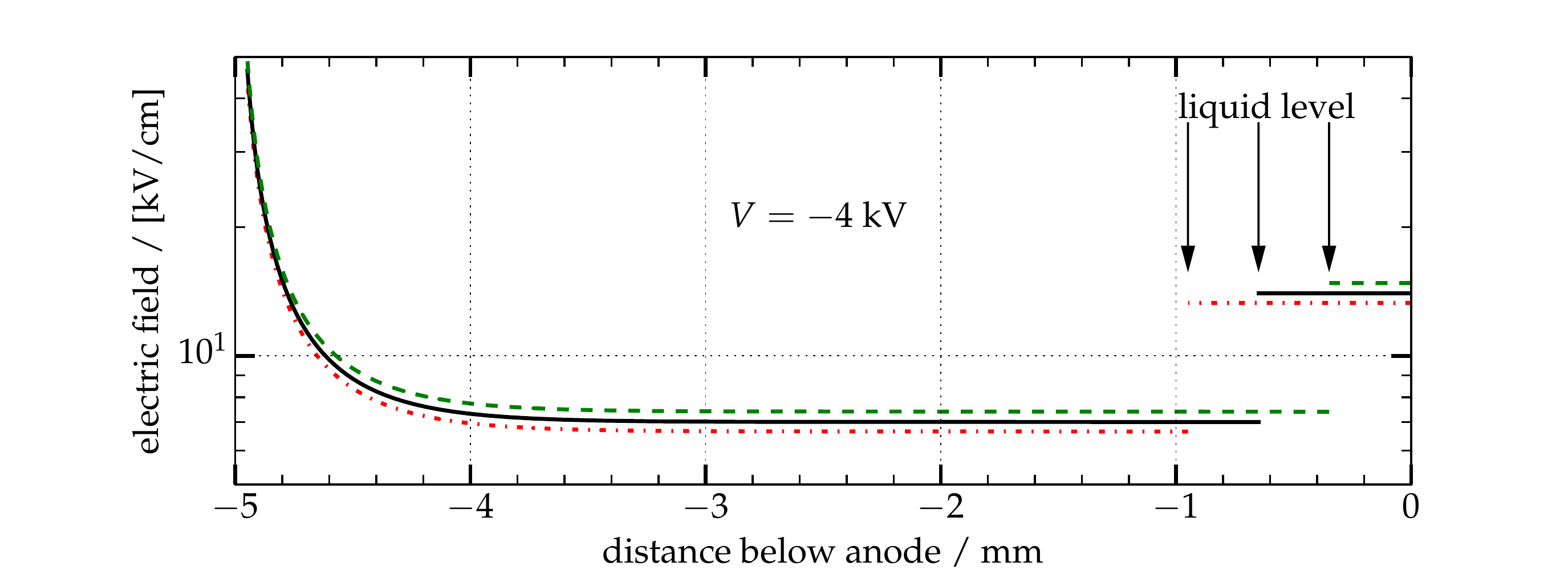}
\vskip -0.1cm
\caption{Calculated electric field as a function of liquid level.}
\vskip -0.5cm
\label{fig:field}
\end{center}
\end{figure}

\section*{Appendix}
For a fixed distance $d$ between cathode and anode, and a liquid height $h<d$, the electric field applied in the liquid is given by 
$E_l = {\epsilon_g V}/({\epsilon_l d + (\epsilon_g-\epsilon_l) h})$,
in which $\epsilon_l=2$ and $\epsilon_g=1$ are the long-wavelength limit of the dielectric constants of the liquid and gaseous states. As required by continuity the electric field in the gas is $E_g = \epsilon_l E_l$. Near the cathode grid wires, the electric field increases rapidly in strength. The field profile is analytical \cite{BlumRolandi:2008}, but cumbersome to write. Fig. \ref{fig:field} shows the field profile in the case $V=4$~kV applied to the cathode, directly above one of the cathode wires. 

There is some width to the distribution of drift times for alpha particles leaving the cathode, and this particular case is useful for comparison against those events with the shortest (cathode-consistent) drift time. The liquid level was estimated from the profile of $E_l$, along with the measured electron drift time and known electron drift velocity in liquid xenon \cite{Gushchin:1982b}. Uncertainty in this procedure is estimated to result in a systematic uncertainty of $\pm0.4$~kV/cm, as indicated in the figures.

\section*{Acknowledgments}
The tunneling hypothesis was suggested by D.N. McKinsey. This work was supported by the U.S. Department of Energy, Office of Science, Office of High Energy Physics, under award number DE-AC02-05CH1123.





\begin{thebibliography}{99}

\bibitem{Bolozdynya:1995} A.I.~Bolozdynya {\it et. al.}, IEEE Trans. Nucl. Sci. {\bf42} 565 (1995).

\bibitem{Aprile:2009dv} 
  E.~Aprile and T.~Doke,
  Rev.\ Mod.\ Phys.\  {\bf 82}, 2053 (2010)
  doi:10.1103/RevModPhys.82.2053
  [arXiv:0910.4956 [physics.ins-det]].

\bibitem{Bolozdynya:2010} A. Bolozdynya, \emph{``Emission detectors,''} World Scientific, Singapore (2010).
  
\bibitem{Gushchin:1982a} E.M. Gushchin, A.A. Kruglov and I.M. Obodovskii, Sov. Phys. JETP {\bf 55} 5 (1982).

\bibitem{Bolozdynya:1999} A.I. Bolozdynya, Nucl. Instr. Meth. A {\bf 422} 314 (1999).

\bibitem{Angle:2011th} 
  J.~Angle {\it et al.} [XENON10 Collaboration],
  Phys.\ Rev.\ Lett.\  {\bf 107}, 051301 (2011)
  Erratum: [Phys.\ Rev.\ Lett.\  {\bf 110}, 249901 (2013)]
  [arXiv:1104.3088 [astro-ph.CO]].

\bibitem{Santos:2011ju} 
  E.~Santos {\it et al.} [ZEPLIN-III Collaboration],
  JHEP {\bf 1112}, 115 (2011)
  doi:10.1007/JHEP12(2011)115
  [arXiv:1110.3056 [physics.ins-det]].

\bibitem{Akimov:2016} D.Yu. Akimov et al,  JINST 11 C03007 (2016).

\bibitem{Xu:2016} J. Xu  {\it et. al.} [LUX Collaboration], \emph{``An investigation of the background electron emissions in the LUX detector,''} APS April Meeting (2016).

\bibitem{Akerib:2016vxi} 
  D.~S.~Akerib {\it et al.} [LUX Collaboration],
  Phys.\ Rev.\ Lett.\  {\bf 118}, no. 2, 021303 (2017)
  doi:10.1103/PhysRevLett.118.021303
  [arXiv:1608.07648 [astro-ph.CO]].
  
\bibitem{Aprile:2017iyp} 
  E.~Aprile {\it et al.} [XENON Collaboration],
  Phys.\ Rev.\ Lett.\  {\bf 119}, no. 18, 181301 (2017)
  doi:10.1103/PhysRevLett.119.181301
  [arXiv:1705.06655 [astro-ph.CO]].

\bibitem{Aprile:2016wwo} 
  E.~Aprile {\it et al.} [XENON Collaboration],
  Phys.\ Rev.\ D {\bf 94}, no. 9, 092001 (2016)
  Erratum: [Phys.\ Rev.\ D {\bf 95}, no. 5, 059901 (2017)]
  doi:10.1103/PhysRevD.94.092001, 10.1103/PhysRevD.95.059901
  [arXiv:1605.06262 [astro-ph.CO]].
  
\bibitem{Essig:2011nj} 
  R.~Essig, J.~Mardon and T.~Volansky,
  Phys.\ Rev.\ D {\bf 85}, 076007 (2012)
  doi:10.1103/PhysRevD.85.076007
  [arXiv:1108.5383 [hep-ph]].

\bibitem{Essig:2012yx} 
  R.~Essig, A.~Manalaysay, J.~Mardon, P.~Sorensen and T.~Volansky,
  Phys.\ Rev.\ Lett.\  {\bf 109}, 021301 (2012)
  doi:10.1103/PhysRevLett.109.021301
  [arXiv:1206.2644 [astro-ph.CO]].
    
\bibitem{Essig:2017kqs} 
  R.~Essig, T.~Volansky and T.~T.~Yu,
  Phys.\ Rev.\ D {\bf 96}, no. 4, 043017 (2017)
  doi:10.1103/PhysRevD.96.043017
  [arXiv:1703.00910 [hep-ph]].
  
\bibitem{Battaglieri:2017aum} 
  M.~Battaglieri {\it et al.},
  arXiv:1707.04591 [hep-ph].

\bibitem{Figgins:1981} P.E.~Figgins, ``The Radiochemistry of Polonium,'' National Academy of Sciences NAS-NS 3037 (1961).

\bibitem{Edwards:2017emx} 
  B.~N.~V.~Edwards {\it et al.},
  arXiv:1710.11032 [physics.ins-det].

\bibitem{Gachkovskii:1970} V.F.~Gachkovskii, T.A.~Kudryavtseva and N.M.~Chirkov, Bulletin of the Academy of Sciences of the USSR, vol. 18 no. 11 (1969). 

\bibitem{Khatipov:2011} S.A.~Khatipov {\it et al.}, Nucl. Instr. Meth. B {\bf 269} 2600 (2011).

\bibitem{Jortner:1965}
J. Jortner, L. Meyer, S.A. Rice and E.G. Wilson, J. Chem Phys {\bf 42} 4250 (1965).

\bibitem{Kubota:1978}
S. Kubota, M. Hishida and J. Raun(Gen),  
J. Phys. C: Solid State Phys. {\bf 11} 2645 (1978).


\bibitem{Ermler:1978}
W.C. Ermler, Y.S. Lee, K.S. Pitzer and N.W. Winter, J. Chem. Phys. {\bf 69} 976 (1978).

\bibitem{Sorensen:2017ymt} 
  P.~Sorensen,
  arXiv:1702.04805 [physics.ins-det].
  
\bibitem{BlumRolandi:2008}
W. Blum, W. Riegler and L. Rolandi (2008) Particle Detection with Drift Chambers, Springer-Verlag, Berlin, Heidelberg

\bibitem{Gushchin:1982b} 
E.M. Gushchin, A.A. Kruglov and I.M. Obodovskii, Sov. Phys. JETP {\bf 55} 4 (1982).


\end{thebibliography}
\end{document}